# The Microstructure of Tool Steel after Low Temperature Ion Nitriding

L. F. Zagonel[a], E. J. Mittemeijer[b], and F. Alvarez[a]

[a] Instituto de Física "Gleb Wataghin", Universidade Estadual de Campinas, Unicamp,
P.O. Box 6165  Campinas, SP, 13083-970, Brazil

[b] Max Planck Institute for Metals Research,
Heisenbergstrasse. 3, D-70569 Stuttgart, Germany

Abstract

The microstructural development in H13 tool steel upon nitriding by an ion beam process was investigated. The nitriding experiments were performed at a relatively low temperature of about 400 ºC and at constant ion beam energy (400 eV) of different doses in a high-vacuum preparation chamber; the ion source was fed with high purity nitrogen gas. The specimens were characterized by X-ray photoelectron spectroscopy (XPS), electron probe microanalysis (EPMA), scanning and transmission electron microscopy (SEM and TEM), and grazing incidence and Bragg-Brentano X-ray diffractometry. In particular, the influence of the nitrogen surface concentration on the development of the nitrogen concentration-depth profile and the possible precipitation of alloying element nitrides were discussed.

Keywords: ion beam nitriding, surface nitrogen concentration, phase transformation, precipitation.

# 1. Introduction

Nitriding is a thermo-chemical process which can lead to pronounced improvements of the mechanical and chemical properties, as the fatigue endurance and the corrosion resistance of steels. [1,2] The process basically consists of nitrogen incorporation through the surface and its subsequent diffusion into the material bulk. Even though nitriding processes are widely used in industrial applications, many fundamental aspects of the process are still not fully understood. Among these, one can single out the role of the (possibly varying with treatment time) nitrogen concentration at the material surface and the nitride precipitation kinetics of iron and alloying elements such as chromium, molybdenum and vanadium, which profoundly influence the nitriding "effect". [3,4,5]

If local equilibrium prevails at the surface, the nitrogen concentration in the solid at the surface is determined by the chemical potential of nitrogen in the nitriding atmosphere. This thermodynamic parameter is crucial for (control of) the efficiency of the nitriding process, also under non-equilibrium conditions. In the gaseous nitriding process, at a non-beginning stage of nitriding, the N concentration at the material surface may indeed be given by the chemical potential of nitrogen in the gas atmosphere. [3,6]

Ion nitriding is a process where such equilibrium conditions at the surface are not expected to occur. Ionic nitriding bombardment takes place at energies larger than the bonding energies of the atoms in the solid to be nitrided, thereby breaking bonds, and knocking-in and sputtering atoms away, emitting secondary electrons, exciting atoms, inducing dislocations, phonons, heat spike, texturization, etc., can occur. This (incomplete) list shows the complexity of the ion nitriding process. Thus, systematic and extensive experimental and theoretical investigation of ion nitriding is imperative in order to control key parameters as the nitrogen concentration at the material surface.

This study contributes to achieve understanding of the effect of nitriding at low temperature (400 ºC; usual nitriding temperatures with well-known, other nitriding processes, as gaseous nitriding, are in the range 500 ºC - 600 ºC); attention is especially paid to the nitrogen concentration at the material surface, and the effects of ion current. The experiments were realized by using a broad Kaufman ion source at constant ion irradiation energy. The Kaufman cell allows a fine control of the beam species, energies and current applied to the sample in a virtually oxygen free nitriding process. [7]

# 2. Experimental procedures

Rectangular samples (2x15x20mm³) were cut from a single tempered AISI-H13 block quenched in air from 1025 °C and tempered at 580 °C for two hours. This heat treatment causes partial precipitation of alloying elements as carbides and sets the bulk hardness to 7.5±0.4 GPa. Table 1 displays the material composition as determined by chemical analysis. The samples were polished up to 1 μm diamond paste and cleaned in an acetone ultrasonic bath. One by one, the samples were introduced into the high vacuum (< $10^{-5}$ Pa) system equipped with the Kaufman cell for ion beam nitriding. The deposition system is attached to an ultra high vacuum chamber (UHV) system (< $10^{-7}$ Pa) for X-ray photoemission spectroscopy (XPS). Thus, immediately after irradiation, the samples can be transferred to the UHV system without atmospheric contamination. The nitriding times and ion current densities used in the experiments are displayed in Table 2. The ions from the Kaufman cell were fixed at 400 eV. The ion source was fed with 8 sccm (standard cubic centimetre per minute, i.e. 8 cubic centimetres at 1 atmospheric pressure) of high purity nitrogen gas (99.999%). A constant pressure of ~ $10^{-2}$ Pa was maintained in the preparation chamber during the nitriding. The ion current densities were determined by measuring the collected ions by a Faraday cup. The sample temperature was maintained at (400±5)°C and the distance between the ion source and the sample was ~25 cm. More experimental details of the implantation system can be found elsewhere.[8]

To determine the elemental surface composition just after the ion implantation, XPS measurements were performed in a UHV chamber attached to the nitriding chamber by a VG-CLAM-2 electron energy analyzer using a non-monochromatic Al Kα radiation and following the procedure described by Hüfner.[9] The photoemission cross-sections were taken as given by Lindau, the elastic means free path is considered to be proportional to $E^{-0.29}$, and the electron energy analyzer transmission is considered proportional to $E^{-1}$.[10] Further material constituents chemical states analysis were performed by ex situ XPS experiments after $Ar^+$ sputtering cleaning (VG Thetaprobe system using monochromatic Al Kα radiation).

Samples cross-sections were analyzed *ex situ* by electron probe microanalysis (EPMA) employing a Cameca SX100 apparatus using a 10kV focused electron beam with an electron current of 100 nA. For this study, the samples cross sections were covered, before EPMA, by a protective electroplated nickel layer and polished (last step 0.25 µm diamond paste). The composition of the nitrided samples was determined from the emitted characteristic X-ray intensities by their comparison with the corresponding intensities recorded from appropriated standards such as pure metals and compounds ($Fe_3C$, $Fe_4N$). The final element content was calculated applying the Φ(θz) method.[11] The estimated error from the EPMA measurements is lower than 0.5at % for all reported results. The morphology of the nitrided microstructure as revealed in specimen cross-sections was investigated by Scanning Electron Microscopy (SEM) using a Jeol JMS-5900LV apparatus. To probe the nanometric precipitates, a Carl Zeiss CEM 902 Transmission Electron Microscope (TEM) operating at 80 kV was employed. The TEM is equipped with a Castaing-Henry energy filter spectrometer for electron energy loss spectrometry (EELS) and a Proscan high-speed slow-scan CCD camera. Suitable TEM samples were prepared in planar view by mechanical polishing and dimpling to the depth of interest, followed by ion milling with 4keV $Ar^+$ ions at an angle of 6°. Crystalline phase analysis was obtained by grazing incidence X-ray diffraction using monocromatized Cu Kα radiation with an incidence angle of 3º (corresponding to an effective penetration/information depth of ~0.1 µm). A larger effective penetration/information depth (~1.4µm) was obtained by using a standard Bragg-Brentano diffraction geometry also using monocromatized Cu Kα radiation (for definition of penetration/information depth, see Ref.[12]).

## 3. Results and discussion

### *3.1 Nitrogen surface concentration; the sticking probability*

For the X-ray photon energy used in the XPS experiments, the information obtained from this technique originates from depths under the surface up to ~ 3 nm.[13] The surface concentration of relevant elements for the investigated specimens is shown in Fig. 1a as function of the ion current density and for constant irradiation time (300 minutes). A plateau level for the element concentrations is observed for ion current densities higher than a characteristic value. By irradiating the sample with a fixed ion current density (1.5 mAcm$^{-2}$), the nitrogen concentration at the surface reaches a plateau before 75 minutes (see Fig. 1b).[14,15,16] This nitrogen concentration corresponds to a stationary state, i.e. a state where a balance is reached between incoming nitrogen and nitrogen losses by self-sputtering or bulk diffusion. The presence of epsilon phase nuclei forming a thin layer could be a possible interpretation of the experimental results. However, the diffractograms corresponding to samples 4 and 6 do not shown any trace of epsilon phase nuclei (see section 3.2). Therefore, one can conclude that the epsilon phase is formed in the stationary state after the accumulation of a relatively high nitrogen concentration.

For completeness, in Fig. 1 the concentrations obtained for a un-nitrided sample are also shown: a virtually oxygen free sample representing the material bulk (data at ion current and time equal to zero, respectively). Usually a thin native oxide layer is always present at the sample surface. However, Fig.1 shows that nitrogen ions remove oxygen by physical/chemical sputtering mechanism.[17]

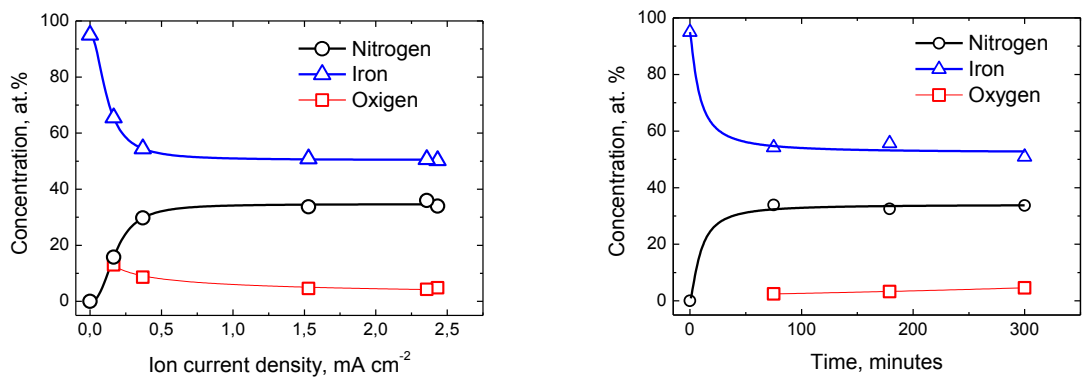

Figure 1: Surface concentrations of nitrogen, iron and oxygen (a) samples nitrided during 300 minutes at different ion current densities and (b) samples nitrided with constant ion current density of 1.5 mAcm$^{-2}$ at different times. The solid lines are guides to the eyes.

The nitrogen sticking probability, i.e., the probability of the nitrogen-containing ions to remain attached to the surface upon bombardment, depends on kinetic ion-surface interactions, backscattering, sputtering, and nitrogen compound desorption. In studies performed by Rabalais *et al.*, concerning $N_2^+$ ion beam bombardment of a pure iron target, the sticking factor was determined as ~ 0.11 at 400 eV.[18,19] This rather low sticking probability indicates that, at the studied ion energies, it is unlikely that the ion upon hitting the surface can loose all its kinetic energy without been backscattered to the vacuum.[20] For the present experiments, the number of nitrogen atoms retained (i.e. incorporated into the sample) per unit of charge may also be smaller than two (atoms per electronic unit charge), which implies that not only $N_2^+$ molecules arrive at the sample surface, but also $N^+$, $N°$, $N_2°$, etc.[21,22] It is noted that the release of nitrogen from the iron nitride surface, which strongly depends on temperature, have been studied by thermal desorption spectroscopy and is not important at 400ºC for iron surfaces.[23]

Assuming the sticking probability to be constant during the nitriding process, the sticking probability can be estimated from the ratio of the ion dose and the total nitrogen incorporated in the material.[24] The assumption of a constant sticking probability is valid provided that the nitrogen concentration at the surface is low, i.e. no significant nitrogen self-sputtering takes place. The total amount of nitrogen taken up by the specimen can be estimated by integrating the nitrogen concentration-depth profile (cf. Fig 5) and assuming that the density remains close to the value of the original material. Sample 4 is particularly suitable for the estimation of the sticking probability since no iron nitrides were formed (Fig. 3) and the measured surface nitrogen concentration is below ~6 at. % (i.e. self-sputtering was not significant). The total nitrogen content in sample 4 (diffusion zone of thickness 10 μm) is ~$3.0 \times 10^{18}$ at cm$^{-2}$. Therefore the sticking probability is ~ 0.15 (nitrogen atoms per electronic unit charge arriving at the sample surface). These quantities have been listed in Table 2 for all the samples investigated.

The sticking probability is an important parameter for understanding the evolution of the nitrogen surface concentration eventually leading to the stationary state at the surface. Indeed a stationary state at the surface of the sample has been reached after a process time which depends on the ion current density (see Fig. 1). Higher sticking probabilities would decrease both time and ion current density needed to reach a stationary state at the surface. After a stationary state at the surface has been achieved, the nitriding process becomes diffusion controlled, i.e. the ionic current does no longer play an important role.[25]

## 3.2 Effect of sputtering on the ion nitriding process: specimen thickness

The impact of energetic nitrogen ions influences and modifies the surface by physical and chemical sputtering, adsorption, and chemical reactions. Sputtering involves loss of material from the specimen: decrease of specimen thickness. On the other hand, nitrogen uptake during the ion beam nitriding can lead to an increase of the specimen thickness as a consequence of stress development and phase transitions. Depending on the experimental conditions, the specimen thickness can either increase or decrease, as experimentally shown by profilometry sweepings across the nitrided and the (properly protected during nitriding) un-nitrided parts of the specimen surface. The height of the step created by the irradiation process as a function of the ion current density and nitriding time is shown in Fig.2. The error bars approximately correspond to twice the surface roughness, which is mainly induced by the nitriding process but is low at such process temperature.[26]

It follows (Fig. 2a) that, up to ~0.5 mAcm$^{-2}$ the nitrided part of the specimen is thicker than the unnitrided part, which is in its original state. At higher ion current densities, sputtering dominates and the surface is eroded, thereby decreasing the specimen thickness. The same is exhibited in Fig.2(b) showing the evolution of the specimen thickness, at a constant ion current density of 1.5 mAcm$^{-2}$, as a function of irradiation time. This plot shows that erosion dominates already before ~75 minutes irradiation time. Roughly linear relations can be adopted for the dependences of the specimen thickness on sputtering time and on ion current density (Figs. 2a and b), after a stationary state for the surface composition has been reached (Fig.1), say, after 40 minutes (see Figs. 1b and 2b ). The total eroded depth, vertical axis in Fig.2, is much smaller then the nitrided case depth. Therefore, the error introduced by the sputtering erosion on the depth profiles as shown in Fig.5 can be neglected.

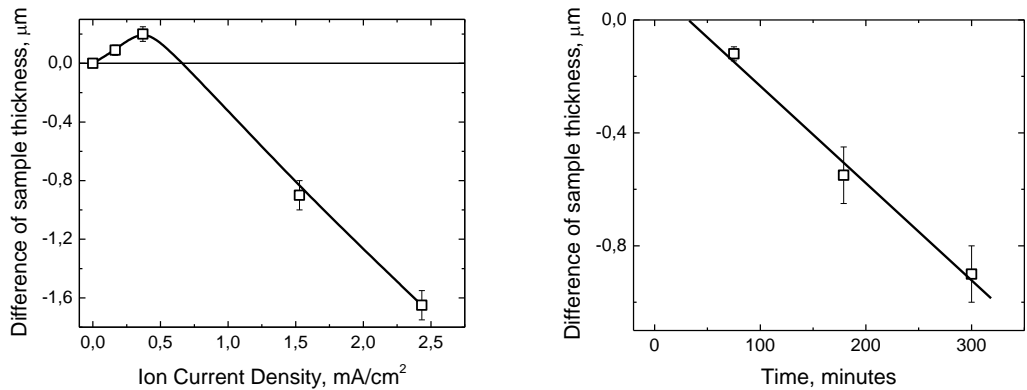

Figure 2: Height difference between nitrided and un-nitrided parts of the specimen for (a) specimens nitrided at constant time (300 minutes) and different ion current densities; and (b) for specimens nitrided at constant ion current density (1.5 mAcm$^{-2}$) and different times. The solid lines are guides to the eyes.

The surface composition and structure influence the sputtering yield. Therefore, in the nitriding process the native oxide layer is expected to have an effect on the sputtering yield. As observed, the erosion caused by nitrogen ions after distinct nitrogen concentrations have been reached is higher than the one occurring for the initial material. For samples 1, 2, 5 and 6, the sputtering rate can be estimated by the ratio between the eroded length (from Fig.2) and the implanted dose (from Table 2), giving approximately 0.05 atoms sputtered away per incident ion.

The initial thickness increment, shown in Fig. 2a, is explained by the development of residual stress upon nitriding. The $\alpha$-Fe 110 reflection (specimen 4 in Fig.3) is shifted with respect to the unnitrided sample. This effect could be interpreted as due to the internal compressive macrostress parallel to the surface induced by the volume expansion in the nitrided zone due to the presence of nitrogen trapped as alloying-element nitride precipitates (Fig.4). The dimension of the 110 lattice planes determined in the direction perpendicular to the surface for the ferrite matrix in Bragg-Brentano geometry (i.e. with the diffraction vector perpendicular to the specimen surface) are 292.1 pm and 288.5 pm for specimen 4 in the nitrided condition and the unnitrided condition, respectively. Assuming a ferrite-lattice expansion proportional to the nitrogen concentration and using the nitrogen profile given in Fig. 5, an increase of specimen thickness of 75±5 nm is estimated. This value for the increase of specimen thickness agrees well with the experimental thickness difference between nitrided and non-nitrided sample regions of 90±30 nm (Fig. 2a).

### *3.3 Crystalline phase analysis of the nitrided surface; iron-nitride formation at the surface*

Grazing-incidence X-ray diffractograms recorded from the surface of specimens investigated are shown in Fig.3. Specimens 1 to 3 gave rise to diffraction peaks indicative of the presence of the $\varepsilon$-Fe$_{2-3}$N phase, i.e. an iron-nitride phase of a relatively high nitrogen concentration. At the lowest ion current density studied (specimen 4) a peak due to ferrite appears, compatible with a low "nitriding potency" of the ion nitriding process under such conditions. Specimen 5 shows that at intermediate nitriding times, iron nitrides are formed. For the lowest studied nitriding time (specimen 6) ferrite appears to be the only phase at the surface. This observation need not be in contradiction with the higher nitrogen concentration obtained from XPS, a technique probing only the first ~3 nm, whereas the grazing-incidence diffractograms shown in Fig. 3 correspond to a penetration/information depth of the order of about 0.1 μm (see section 2). Diffractograms recorded in Bragg-Brentano geometry (compatible with penetration/information depths of about 1.4 μm) indicate that the layer containing $\varepsilon$-$Fe_{2-3}N$ phase is thin (≤1 μm) for all studied samples, as expected for the low process temperature applied (400°C).

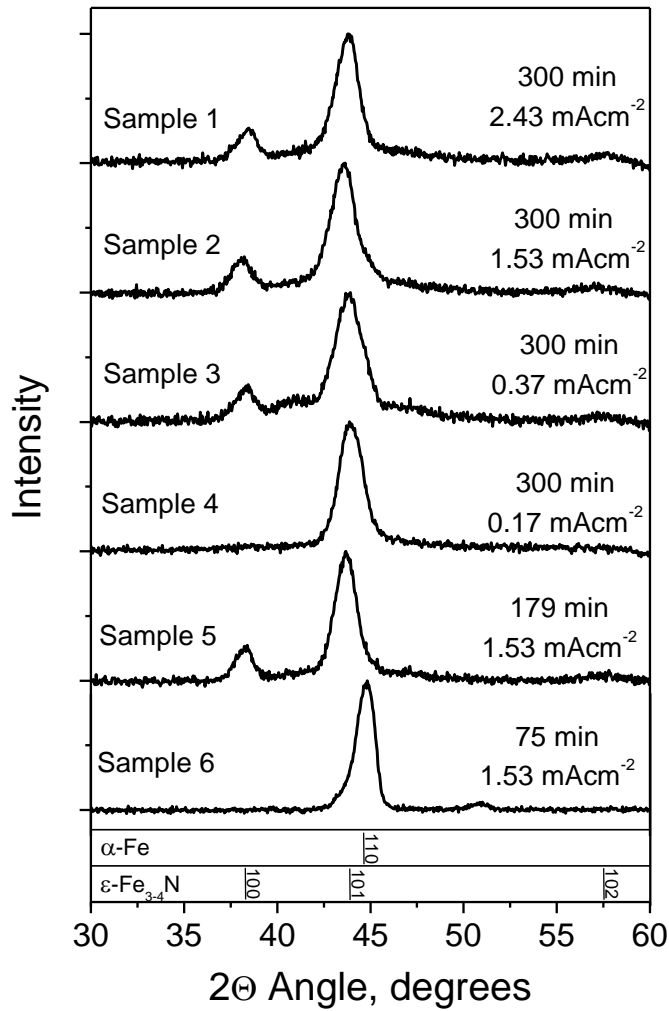

Figure 3: Small angle X-ray diffractograms recorded from specimens nitrided under different conditions, as indicated.

### 3.4 *Chemical state of the alloying elements; alloying element nitride formation*

Besides providing values for the concentrations of the various elements in the surface region of the analysed specimen, XPS gives information on the chemical state of these elements. This information is obtained by analysing the peak positions (binding energy values) associated with the studied element core level electrons. The XPS spectra of nitrided samples show that the positions and shapes of the bands associated with chromium, molybdenum, and vanadium differ from those observed for an untreated clean (Ar+ sputtered) reference sample. As observed in Figs. 4 (b) and (c), the bands associated with vanadium and molybdenum electron core levels shift towards high binding energies for samples 2 and 3 ($\Delta E=+0.2$ eV). However, for sample 4 the result is similar to the one found for the unnitrided reference sample. The band associated with chromium, on the other hand, displays only a slight increase of the high-binding energy tail (samples 2 and 3). The absolute amount of the alloying elements was similar for all samples, including those with an iron nitride phase at the surface.

These shifts in binding energy are ascribed to the formation of alloying element nitrides, occurring as precipitates in the nitrided matrix, such as CrN, VN, and MoN. The results indicate that up to an uptake corresponding to ~5at.% nitrogen, the binding energies of the alloying elements remain unchanged and therefore precipitation does not occur substantially up to this level of nitrogen uptake (see Fig. 7). For higher nitrogen concentration values, as holds for samples 2 and 3, the 2p levels of V and Mo reveal a different chemical state for V and Mo, which is considered to be indicative for the presence of alloying element nitride precipitates. On the other hand, the Cr 2p level spectra indicate that chromium carbides are present in the virginal material and that these are not dissolved upon nitriding to form chromium nitride (see section 3.5).

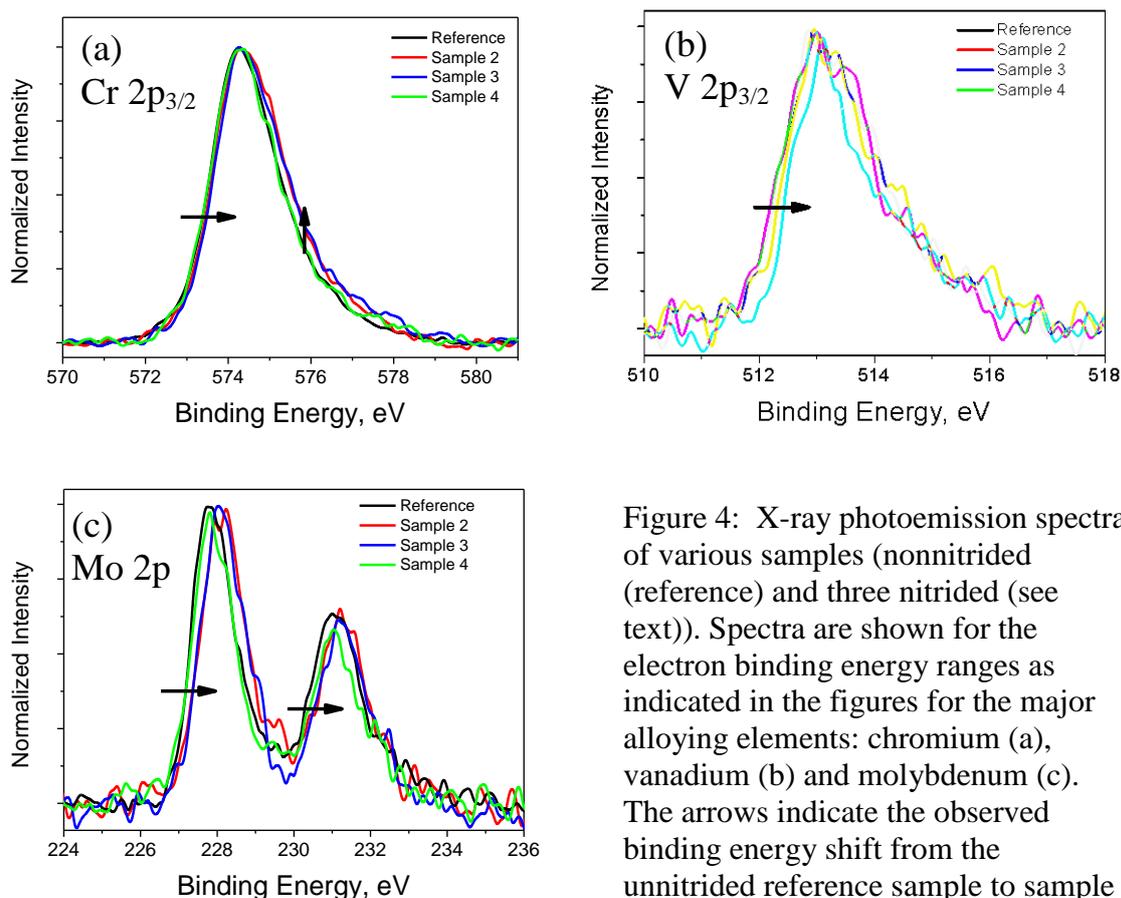

Figure 4: X-ray photoemission spectra of various samples (nonnitrided (reference) and three nitrided (see text)). Spectra are shown for the electron binding energy ranges as indicated in the figures for the major alloying elements: chromium (a), vanadium (b) and molybdenum (c). The arrows indicate the observed binding energy shift from the unnitrided reference sample to sample 2 (high nitrogen content).

## *3.5  Nitrogen concentration-depth profiles; precipitate particles in the nitrided zone*

Nitrogen concentration-depth profiles as determined by EPMA from sample cross sections (see section 2) are shown in Fig.5 for:  a) specimens of the same nitriding time and treated at different ion current densities (Fig. 5a) and, b) for specimens treated at constant ion current density for different nitriding times (Fig.5b).

Fig.5(a) shows that the nitrogen penetration depth is about the same for all samples. In first order approximation this may be expected for equal nitriding time treatments, if the same concentration of dissolved nitrogen would occur at the surface, which however is not truly the case (see sections 3.1 and 3.4), and if solid state diffusion of nitrogen would be the dominantly rate controlling process. The amount of nitrogen in the sample increases monotonically from samples 4 to 1, i.e., on increasing ion nitrogen current density. This behaviour is ascribed to the earlier attainment of a stationary state (cf. section 3.1) at the sample surface. The  nitrogen profiles observed for samples 1 and 2 are more or less similar and this suggests that the ~ 60%  increase of the ion-beam current applied for sample 1 with respect to sample 2 (cf. Table 2) does not produce pronounced changes of the nitrogen depth profile. Indeed, after the establishment at the surface of a stationary state, as appears to hold for samples 1 and 2 (see Fig.5a), the magnitude of the ion-beam current is not important for the progress of nitriding in the solid substrate. In the case of sample 4a a stationary state at the surface has evidently not been realized (Fig. 1 and Fig. 3). Fig.5 (b) displays the effect of nitriding time for samples with identical (final) nitrogen surface concentrations, as follows from Fig. 1(b). Sample 6 has a shallow nitrogen profile as a consequence of the short nitriding time even though the surface stationary state concentration has already been established.

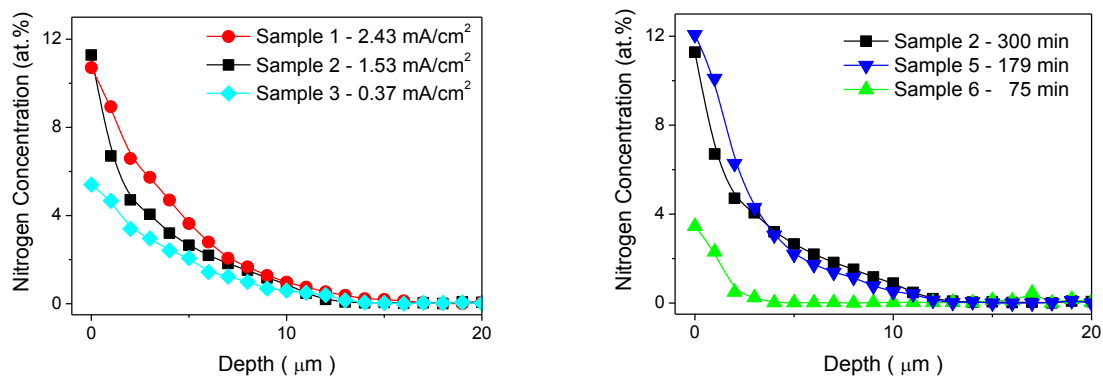

Figure 5: Nitrogen concentration-depth profiles (EPMA results) (a) for specimens nitrided for 300 minutes, for variable ion current density and (b) for specimens nitrided at constant ion current density (1.5 mAcm$^{-2}$) for variable nitriding time. Lines shown are guides to the eye.

Near to the surface, the nitrogen concentrations as obtained from EPMA in the specimen cross section are smaller than those determined by XPS (Fig.1). This is expected, since XPS, applied to the sample surface, and EPMA, applied to the sample cross section, probe up to a depth of ~ 3-5 nm and up to a volume of diameter about 1 μm, respectively. Therefore, EPMA results are quantitatively more reliable. However, it should be realized that EPMA, on the sample cross section, cannot be performed very close to the surface, because of the size of the analysed specimen volume by EPMA in a single measurement, whereas XPS applied to the surface of the nitrided sample provides data pertaining to the very surface adjacent region (see small depth of analysis indicated above). In particular for sample 6, the rapid decrease of the nitrogen concentration across the first micrometers leads to a considerable different between the EPMA and XPS nitrogen concentration values.

The highest value found for the concentration of nitrogen in the diffusion zone (substrate with ferritic matrix) is about 10 at.%, recognizing that the iron-nitride layer at the surface has a thickness less than about 1 µm (cf. section 3.3); see Fig.5 and Fig.3. Because the solubility for nitrogen dissolved in the ferrite matrix is much less (even considering the possibility of dissolved "excess" nitrogen (see Ref.27), this concentration value also suggests that the predominant part of the nitrogen must be present as alloying element nitride precipitates. The possible presence of excess nitrogen, due to the misfit-strain fields surrounding the tiny, coherent, alloying element nitride particles,[28,29] may be supported by the difference on the amount of nitrogen present (10at.%) and the amount of nitrogen that could be combined to form alloying elements precipitates (maximum of about 5at.%). The nitrogen concentration difference, about 5at.%, would be present as excess nitrogen, even if a part of these atoms could be forming fine iron nitride precipitates. Nevertheless, the latest were not observed by XRD, ruling out this possibility.

The presence of fine nitrides particles was confirmed by TEM analysis. Two electron transparent foils were prepared from the nitrided sample 2. The first specimen was taken at ~5-10µm depth, i.e. within the diffusion zone/nitrided region; the second specimen was taken at depth ~50µm, i.e. in the nonnitrided core region of the sample. In the former specimen, the TEM and EELS measurements revealed only the presence of carbide particles (not shown). In particular, TEM analysis of the first specimen demonstrated the presence of vanadium nitride precipitates in the diffusion zone (see Fig.6). Indeed, the EELS spectra of these precipitates showed relatively high concentrations of nitrogen and vanadium.

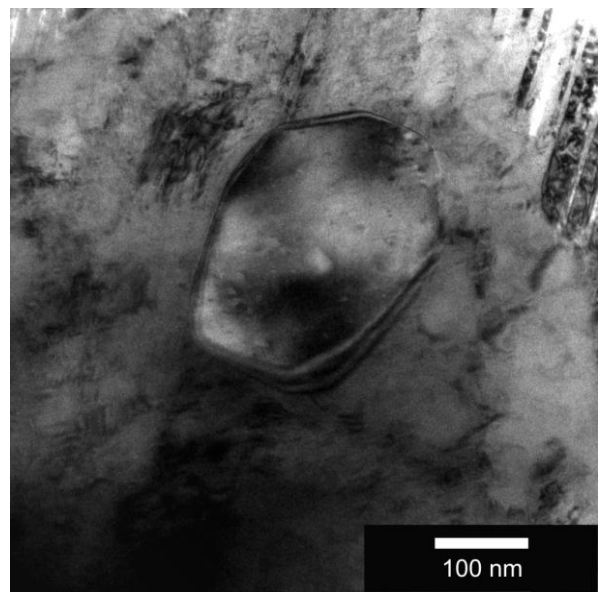

Figure 6: TEM micrograph (bright field) of vanadium-nitride precipitate within the diffusion zone of sample 2.

An EPMA line scan across the nitrided case and into the unnitrided core is shown in Fig. 7 for sample 1. The coincidence of the carbon and chromium concentration peaks suggests that a chromium-carbide particle occurs in the transition region of the nitrided zone and the unnitrided material core. The conversion of chromium carbide into chromium nitride (as observed for gas nitrided Fe-Cr-C alloy [30, 31]) is probably kinetically hindered in the present experiments due to the relatively low nitriding temperatures applied in the current experiments (~400°C). This result is in agreement with the XPS data presented and as discussed in section 3.2 (XPS Cr spectrum).

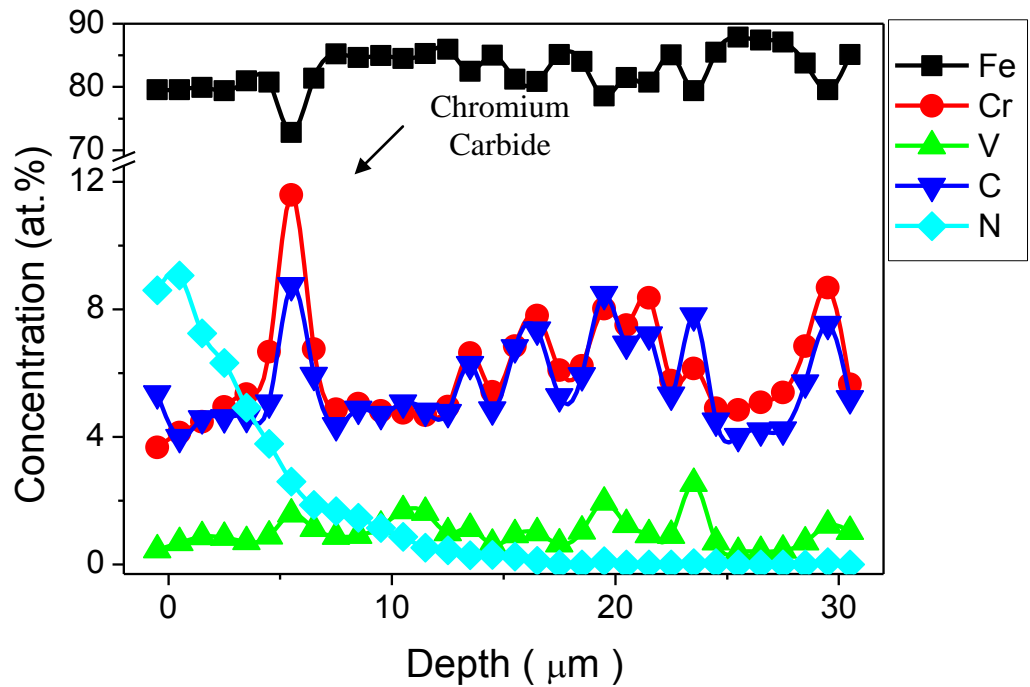

Figure 7: Concentration-depth profiles obtained by an EPMA line scan in the sample 1 cross section. The presence of a chromium-carbide particle is revealed in the transition region of the nitrided case and the unnitrided core material.

## 4. Conclusions

- Upon ion nitriding the nitrogen surface concentration increases gradually until a stationary value has been attained, as the outcome of balancing of the reacting nitrogen ions arriving at the sample surface and the nitrogen removed by self-sputtering and by diffusion into the material bulk. A typical value for the sticking probability is ~ 0.15 N atoms per electronic unit charge.

- Upon ion nitriding the specimen thickness can decrease due to the effect of the accompanying sputtering and the specimen thickness can increase due to the uptake of nitrogen and the associated development of alloying element nitrides and the corresponding occurrence of a compressive macrostress parallel to the surface. At current densities above, say, 1 mA/cm$^2$ the sputtering effect dominates and the emergence of a stationary state for the nitrogen concentration at the surface (see above) parallels the occurrence of roughly linear relations between increase of treatment time and decrease of specimen thickness, and increase of current density and decrease of specimen thickness.

- Upon ion nitriding quenched and tempered tool steel at 400 ºC:
(i) at the surface a thin iron-nitride (epsilon nitride) layer of thickness less than 1μm can develop;
(ii) the highest value found for the concentration of nitrogen in the diffusion zone (substrate with ferritic matrix), i.e. near to the interface with the thin epsilon-nitride surface layer, is about 10 at.%. The largest part of this nitrogen is incorporated in alloying element nitride precipitates;
(iii) the alloying elements V and Mo precipitate as tiny nitride particles; the chromium carbides present initially in the matrix are not converted to CrN at this low nitriding temperature, in contrast with nitriding at elevated temperature as with gaseous nitriding treatments.

## Acknowledgments

Part of this work was performed during a stay of LFZ at the Max Planck Institute for Metals Research in Stuttgart, Germany. The authors are grateful to Mrs. S. Haug and Mr. W.-D. Lang from the Max Planck Institute for Metals Research for assistance with the electron probe microanalysis measurements and TEM sample preparation, and to C. A. Piacenti from Unicamp for his technical assistance. This work is part of LFZ´s Ph.D thesis and was partially supported by FAPESP, project # Nº 97/12069-0. FLZ is a FAPESP and DAAD fellow. FA is CNPq fellow.

Tables:

Table 1: AISI H13 steel composition; data in atomic percent.

| Element | Fe | C | Mn | Si | Cr | Mo | V |
|---|---|---|---|---|---|---|---|
| Content (±0.1) | 87.3 | 2.5 | 0.4 | 2.1 | 5.9 | 0.8 | 1 |

Table 2: Summary of applied ion nitriding parameters. Grey cells highlight constant-time and constant-ion-current-density specimen sets. The nitrogen beam dose, amount of retained nitrogen (i.e. the nitrogen uptake) and calculated nitrogen sticking probability (see section 3.6) are also displayed. All nitriding experiments were performed at 400 ºC.

| Sample | Ion current density (mAcm$^{-2}$) | Irradiation Time (min) | Dose ($\times 10^{19}$ electronic charge units cm$^{-2}$) | Retained nitrogen ($\times 10^{18}$ at. cm$^{-2}$) | Sticking probability (number of N atoms retained per electronic unit charge) |
|---|---|---|---|---|---|
| 1 | 2.43 | 300 | 273 | 4.7 | 0.017 |
| 2 | 1.53 | 300 | 172 | 4.4 | 0.026 |
| 3 | 0.37 | 300 | 4.2 | 3.6 | 0.086 |
| 4 | 0.17 | 300 | 1.9 | 3.0 | 0.155 |
| 5 | 1.53 | 179 | 10.3 | 4.2 | 0.0410 |
| 6 | 1.53 | 75 | 4.3 | 0.71 | 0.0165 |